# Planetary harmonics in the historical Hungarian aurora record (1523-1960)

NICOLA SCAFETTA

ACRIM, Coronado, CA, 92118 USA & Duke University, Durham, NC, 27708 USA

*ns2002@duke.edu*

*nicola.scafetta@gmail.com*

and

RICHARD C. WILLSON

ACRIM, Coronado, CA, 92118 USA

rwillson@acrim.com

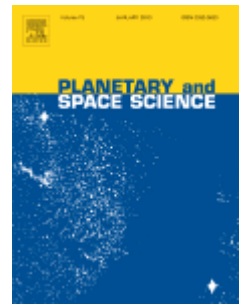

## Abstract

The historical Hungarian auroral record extends from 1523 to 1960 and is longer than the sunspot record. Harmonic analysis reveals four major multidecadal secular cycles forming an approximate harmonic set at periods of 42.85, 57.13, 85.7 and 171.4 years. These four frequencies are very close to the four major heliospheric oscillations relative to the center of mass of the solar system caused by Jupiter, Saturn, Uranus and Neptune. Similar frequencies are found in solar radiation models based on long cosmogenic isotope records (Steinhilber et al. 2012) and in long records of naked-eye sunspot observations (Vaquero et al., 2002). Harmonic regression models are used to reconstruct and forecast aurora and solar activity for the period 1956-2050. The model predicts: (1) the multidecadal solar minimum in the 1970s that is also observed in the sunspot record; (2) a solar maximum in 2000-2002 that is observed in the ACRIM total solar irradiance satellite composite; (3) a prolonged solar minimum centered in the 2030s. These findings support a hypothesis that the Sun, the heliosphere and the terrestrial magnetosphere are partially modulated by planetary gravitational and magnetic forces synchronized to planetary oscillations, as also found in other recent publications (Scafetta, 2010, 2012a, 2012c, 2012d; Abreu et al., 2012; Tan & Cheng, 2012).

# 1 Introduction

Auroras are among the most spectacular events observed in the sky (Akasofu, 2009; Brekke, 2012). The aurora oval covers most of the Arctic and Antarctic regions of the Earth, typically within 10°-20° from the magnetic poles. Auroras are space weather events regulated by the Sun, the heliosphere and the terrestrial magnetosphere (Moldwin, 2008). Auroras caused by the most energetic solar eruptions are occasionally observed at mid-to-low latitudes, facilitating the numerous historical records, in particular from central and southern European countries.

Historical auroral sighting records are valuable because they are the longest direct observational records available for studying solar and space weather dynamics. Catalogs of historical auroral records, in particular from the mid-latitudes, have been compiled since the 19th century (Fritz, 1873; Fritz, 1881; Lovering et al., 1868; Angot, 1897) and updated later by other authors: see Krivsky & Pejml (1988), Silverman (1992) and numerous references therein.

The auroral global catalogs vary greatly in reliability. Before 1700 the data are sparse for various reasons; journal press was uncommon and many older documents may have been lost or destroyed. From 1700 to 1900 numerous records exists, but highly fragmented. Since 1900 the historical naked-eye auroral records have also become sparse for various reasons, including the addition of street lighting that has made it more difficult to observe auroras from central Europe. Since 1950 naked-eye collections of direct auroral sightings have been discontinued in favor of photographic and electronic records.

Reliability suffers as well because the global catalogs mix fragmented records from different regions that cover different time intervals. In some records sightings from North America, the United Kingdom, central and southern Europe and Asia are commingled (Angot, 1897; Křivský & Pejml, 1988). However, the frequency of sightings varies greatly among these regions, as demonstrated by Loomis in 1860 (Akasofu, 2009). The auroral oval moves as well;

so in time the frequency of auroral sightings may increase in one region and decrease in another. When inhomogeneous records are combined into a single comprehensive catalog spurious trends and artificial patterns may emerge, as evident in Křivský & Pejml (1988): see also Figure 1B.

Misinterpretations due to inhomogeneity artifacts in the global records can be minimized by studying carefully collected records as lengthy as possible from a well-defined region. Records from regions where auroral sightings are relatively rare may yield to records statistically more reliable on decadal scales since rare events are more likely to elicit the interest of observers and result in their documentation.

We have chosen to study the historical Hungarian auroral record (Nordlichtbeobachtungen in Ungarn) that extends from 1523 to 1960 (Rethly & Berkes, 1963) to determine whether this direct observational solar related record presents the signature of planetary harmonics recently found in other related solar activity records (Scafetta, 2010, 2012a, 2012c, 2012d; Abreu et al., 2012; Tan & Cheng, 2012). The Hungarian auroral record is chosen because it is one of the longest available. The Hungarian record covers 438 years, longer than the 402-year sunspot number record that began in ~1610 with Galileo's telescope observations. The Hungarian auroral record appears to be sufficiently accurate and stable for our purpose, as explained in Section 2. We compare the auroral analysis results with those referring to a comprehensive planetary orbital function and to a lengthy solar radiation model based on cosmogenic isotope records (Steinhilber et al. 2012).

## 2 Data and analysis

The annual frequency of the historical Hungarian auroral record (Rethly & Berkes, 1963) is reported in Table 1. In this region auroral sightings are rare with, on average, about one event every two years. To highlight the multidecadal dynamical pattern in the record we used an 11-year moving average algorithm since the auroral frequency correlates with the ~11-year solar cycle (Akasofu, 2009) and only a few auroras per year are present. The results are shown in Figure 1.

Figure 1A compares the Hungarian auroral record with the sunspot number record. The maxima of the Hungarian auroral record (near 1610, 1725, 1784, 1834, 1870, 1913 and 1945) correspond closely with the maxima in the sunspot record. The Maunder (1645-1715) and Dalton (1790-1830) solar activity minima are clearly seen in both records. The sunspot number record presents a continuous upward long term trend since 1600 (average before 1780 = 27, average from 1780 to 1960 = 48; ratio 1.77). The Hungarian auroral record presents a statistically compatible trend (average before 1780 = 0.44, average from 1780 to 1960 = 0.71; ratio 1.61). During the sunspot minima it returns to zero because only the most massive solar eruptions can produce auroras visible at low latitudes and these eruptions are very rare during prolonged solar minima.

The statistical compatibility of the sunspot record and aurora record trends before and after 1780 suggests the Hungarian aurora record is sufficiently stable. Thus, the fact that polar aurora might have been more or less systematically observed in Hungary after the Universitäts-Sternwarte in Budapest was established in 1780 does not appear to have significantly changed the statistical properties of the aurora record. The slight increase of the number of Hungarian polar aurorae since 1800 is more likely due to the increased solar activity, as revealed by the sunspot record, than to an artifact due to more regular university-based observations.

Secondary discrepancies exist such as a sunspot peak in ~1955 that does not match the Hungarian auroral record, which shows a peak in ~1945. However, an auroral peak in the 1940s is found also in a compilation of German auroral observations from 1946 to 1964, which shows a peak in the 1940s (Schröder, 2011). Solar activity may actually have peaked in the 1940s, as some solar proxy models predict (Hoyt & Schatten, 1997; Thejll & Lassen, 2000; Scafetta, 2012c). These results seem to indicate that the physics regulating auroral phenomena is coupled with, but not simply related to sunspot occurrence.

Figure 1B compares the Hungarian record with the global mid-latitude aurora catalog (Křivský & Pejml, 1988). A good correlation is observed in the timing of the peaks, although the global catalog presents a strong step-increase artifact in 1700 due to the addition of records from northern regions (average 1523-1700 = 3, average 1700-1900 = 26; ratio 8.6, while the Hungarian record increased by only 1.7 between the same periods). One discrepancy refers to the dominant aurora peak in 1849 that does not appear in the Hungarian catalogue. However, the existence of this 1849 peak is controversial because it is also not seen in the high frequency aurora record for New England (Silverman, 1992).

Corrections to the auroral record, required because of the movement of the Northern magnetic pole (NMP), would change the annual frequency of auroral sightings. These are expected to be quite small, however, and to affect the multisecular scale (>400 yr) only slightly since from 1523 to 1960 the NMP moved within a relatively small region located in north Canada (between 220E-270E and 68N-80N) describing a quasi cyclical pattern (McElhinny and McFadden, 2000): see also Figure 5 in the Appendix.

Note that the geographical extent of Hungary has changed during the last 5 centuries and its present small extent exists since 1918 onward. Before 1918 Hungary was larger, including part of the Balkans, Slovakia and Transylvania (currently in Romania). This relatively small regional change is unlikely to have introduced significant artifacts in the aurora sighting frequency record since auroras can be seen for several hundred kilometers. Figure 1 shows the Hungarian auroral record relatively well correlated with both the sunspot record and the global mid-latitude aurora record, which covers a far greater region than the historical sizes of Hungary.

Figure 2 depicts the normalized Lomb periodogram of the Hungarian record. Note that each region of the Earth is also characterized by a specific auroral sighting probability coefficient that depends on the local climate: however, a simple scaling coefficient does not affect the periodogram because of the normalization procedure. Variations of this coefficient due to multidecadal/secular climate changes are expected to be relatively small, on the order of a few percent since climatic temperature changes are up to ~1 $^o$C since the Little Ice Age and mostly correlate to solar variations (Scafetta and West, 2007; Scafetta, 2009, 2012a). Climatic variations should work as a modest feedback for auroral sighting, which also should not alter the power spectrum patterns. A fluctuating random component of this coefficient contributes to a noise component that is taken into account by evaluating the confidence values of the spectral peaks against white and red noise backgrounds, as shown in Figure 2.

Four major power spectral peaks are observed at about 41-44, 54-60, 80-90, 150-185 years, and appear to form an almost perfect harmonic set at 42.85 yr, 57.13 yr, 85.7 yr, 171.4 yr. The latter periods are, respectively, the fourth (1/4 P), third (1/3 P) and second (1/2 P) harmonics of the fundamental harmonic at P = 171.4 yr corresponding to the conjunction of Uranus ($P_U$ = 84.01 yr) and Neptune ($P_N$ = 164.8 yr). The conjunction period is given by the equation: $P_{UN}=1/(1/P_U-1/P_N)$. These four frequencies are close to four major solar system oscillations related to Jupiter, Saturn, Uranus and Neptune, which are found among the resonance frequencies in the Sun's motion about the center of mass of the solar system (CMSS) (Bucha et al., 1985; Jakubcová & Pick, 1986). Smaller spectral peaks with periods shorter than 40 yr are observed above the 99% confidence level at about 28.5 yr and 35 yr. These peaks correspond to the fifth (1/5 P) and sixth (1/6 P) harmonics of the P = 171.4 yr cycle, and are consistent with the studies of Jakubcová & Pick (1986).

Figure 3A confirms the above claim by comparing two alternative power spectra of the auroral record against the power spectrum of the z-axis coordinate of CMSS about the Sun (equatorial plane). The latter is just one of the convenient observables that can be used to detect the frequencies of the major oscillations induced by the planets. Other functions of the orbits of the planets may be used such as the solar velocity, acceleration, radii of curvature, angular momentum, etc. (Bucha et al., 1985). However, according Taylor's theorem these functions are characterized by common spectral frequencies, only the power associated to each frequency may vary from function to function. Because here we are interested only in determining a generic set of planetary harmonics characterizing the solar system, the particular function of the planetary orbits used for our calculations is not particularly important. Our computations were made using the Jet Propulsion Laboratory HORIZONS ephemeris system from B.C. 3000 to A.D. 3000.

The following major spectral frequency peaks are found: about 43 yr (forth harmonic of Uranus-Neptune synodic cycle); 45.3 yr (Saturn-Uranus synodic cycle); about 57 yr (third harmonic of Uranus-Neptune synodic cycle); about 61 yr (Jupiter-Saturn major tidal beat cycle); about 81-87 yr (second harmonic of Uranus-Neptune synodic cycle, and Uranus orbital cycle); about 155-175 yr (Neptune orbital cycle and its synodic cycle with Uranus). Note that the power spectrum for determining the planetary oscillations uses data for 6000 years, while the Hungarian auroral record covers only 438 years. This may explain why the former presents more details showing additional frequency peaks than the latter.

Solar radiation models based on long cosmogenic isotope records present spectral frequency peaks compatible with planetary deduced harmonics (Scafetta 2012c, figure 4). Abreu et al. (2012) studied power spectra of long solar radiation models and found that several multi-secular spectral peaks correspond to planetary harmonics. Figure 3B shows two alternative power spectra (within the 40-200 yr period band) of the solar model record proposed by Steinhilber et al. (2012) covering 9,400 years during the Holocene. The data are 40-year running means and scales smaller than 80 yr are strongly smoothed out. The figure highlights major spectral frequency peaks at about 175, 86 and 57 yr plus a peak at about 43 yr, which appears quite small likely because the data are 40-year running mean smoothed. These four frequencies are compatible within a 2% uncertainty with those found in the auroral record. Note that the 175 yr cycle falls between the 171.4 Uranus/Neptune synodic cycle and the 178.7 yr solar wobbling cycle (Jose, 1965).

Figure 3B shows additional spectral peaks already linked to planetary motion at periods near 45, 61, 113 and 130 yr (Scafetta, 2012c). Other spectral peaks near 72-75, 91, and 98 and 122 yr are seen in Figure 3A among the planetary oscillations. Restricting the spectral analysis of Steinhilber's model during the auroral period 1520-1960 AD confirms the existence of a major 140-190 yr oscillation compatible with the graph depicted in Figures 2 and 3.

Some of the above results are found in other direct solar observational records. For example, long records of naked-eye sunspot observations present major frequency peak at about 11 yr, 15 yr, 46 yr, 60 yr, 85 yr, 115 yr, and ~250 yr (Vaquero et al., 2002), that are compatible with the spectral peaks depicted in Figures 2 and 3.

Figure 4 shows the Hungarian auroral record, smoothed by an 11-year moving average, and its decomposition into four Fourier band-pass filter components (thick, colored curves) centered around the found four frequencies: $H_1(x)$ captures the dynamical variability within the scale of 130-233 yr; $H_2(x)$, 74-130 yr; $H_3(x)$, 49-74 yr; $H_4(x)$, 35-49 yr. The band-pass filtering of the aurora record was accomplished by first calculating its fast Fourier transform (FFT) with appropriate zero padding to increasing the resolution (Muller and MacDonald, 2000) and then by applying an inverse FFT after having set all FFT coefficients outside the frequency band of interest to zero. The thick gray curve is the sum of the four harmonic components. The four oscillations synchronize to form large peaks around 1610, 1784 and 1945. These are separated by about 171.4 years, which is the synodic period of Uranus and Neptune. The prolonged periods of minima, such as the Maunder Minimum (1645-1715), occur during periods of destructive interference patterns among the harmonics.

The black thin curves in Figure 4 are regression harmonic models of the four auroral Fourier components, each made with three optimal frequencies to better capture the beat modulation of the type:

$$h_i(x) = A_1 \cos[2\pi(t-T_1)/P_1] + A_2 \cos[2\pi(t-T_2)/P_2] + A_3 \cos[2\pi(t-T_3)/P_3], \quad (1)$$

with i=1, 2, 3 and 4. The nine regression coefficients for each of the four curves are reported in Table 2. The four curves $h_i(t)$ are summed and the combined curve is shown as a thin black line extending the auroral record of the top of the figure. The blue curve is a direct regression model of the auroral record using the four auroral frequencies highlighted in Figures 2 and 3: that is, we use the function

$$f(x) = A_1 \cos[2\pi(t-T_1)/P_1] + A_2 \cos[2\pi(t-T_2)/P_2] + A_3 \cos[2\pi(t-T_3)/P_3] + A_4 \cos[2\pi(t-T_4)/P_4] + C, \quad (2)$$

whose 13 regression coefficients are reported in Table 2. The regression models find best harmonics to fit the data. The estimated frequencies are compatible with the theoretical ones estimated via planetary harmonics.

The two alternative regression models predict that the Sun experienced a multidecadal minimum activity during solar cycle 20 (1964-1976), which was concurrent with a significant sunspot cycle minimum (see also Figure 1A), and a multidecadal maximum activity during solar cycle 23, whose maximum occurred in 2000-2002. The two models predict another significant solar activity minimum during the 2030s.

A multidecadal solar activity increase from the 1970s to 2002 and a decrease thereafter conforms to the satellite total solar irradiance record, as constructed in the ACRIM total solar irradiance satellite composite (Willson and Mordvinov, 2003). Note that the same pattern is not recovered by the PMOD total solar irradiance satellite composite which shows a gradual decrease since 1980 to the present (Fröhlich, 2006). Fröhlich's PMOD composite is based on the assumption that some solar irradiance satellite measurements need to be altered to conform to the predictions of solar proxy models, contrary to the opinion of the original satellite experiment teams. Thus, the harmonic model depicted in Figure 4 would contradict the PMOD composite and support the ACRIM composite (Scafetta & Willson2009), implying that the ACRIM composite better captures the solar irradiance behavior during the last three decades.

Our prediction of a solar irradiance peak around 2000 and a significant prolonged solar minimum near 2030 has also been predicted by an alternative solar model based on planetary harmonics (Scafetta, 2012c), which is also compatible with the auroral maximum observed in the 1940s. The observed pattern forms a quasi 60-year cycle that can be associated to the 57-61 yr planetary oscillations.

## 3 Discussion and conclusion

The Hungarian auroral record (1523-1960) is one of the longest auroral sighting records available. It can be interpreted as a frequency index of the most energetic solar eruptions that are capable of causing aurorae observable at the low latitudes in Hungary. A correlation between planetary alignment and the largest solar flare eruptions have been observed (Hung, 2007). Figure 1 shows that the Hungarian record exhibits periods of higher sighting frequency relatively well correlated with other auroral catalogues (Křivský & Pejml, 1988; Silverman, 1992) and with the sunspot record. In the latter case, the observed secondary discrepancies (e.g. the Hungarian auroral record peaks in ~1945 vs. the sunspot number peaks in ~1955) seem to indicate that the physics regulating auroral phenomena is coupled with, but not simply related to sunspot occurrence.

Harmonic analysis reveals that the Hungarian auroral record is characterized by at least four major spectral frequencies that form an almost perfect harmonic scale at periods of about 42.85,

57.13, 85.7, 171.4 years. These frequencies are close to four major resonance frequencies of the orbital alignments of Jupiter, Saturn, Uranus and Neptune, and are part of the harmonics of the basic period of 171.4-178.4 yr that characterizes the solar system (Jakubcová & Pick, 1986).

Planetary orbital harmonics present in auroral records have been the subject of other studies as well (Charvátová-Jakubcová et al., 1988; Scafetta, 2012a). Similar harmonics are detected in solar proxy records (Ogurtsov et al., 2002; Scafetta, 2012c). Here we have confirmed this result by studying the high resolution power spectrum of a solar radiation model based on long cosmogenic isotope records proposed by Steinhilber et al. (2012). Our finding is that Steinhilber's solar proxy model presents numerous planetary harmonics, extending the findings of Scafetta (2012c) and Abreu et al. (2012).

One important result is that the adoption of the longer Hungarian aurora record makes it possible to distinguish the quasi 60 yr and 85 yr oscillations commonly observed in solar proxy records (Ogurtsov et al., 2002; Scafetta, 2012c). This operation was not possible using the short mid-latitude aurora catalog (Křivský & Pejml, 1988) from 1700 to 1900 (Scafetta 2012a) where a modulation of the mid-latitude aurora global catalog was found to be well described by a quasi 60-year oscillation. In the Hungarian record, however, the quasi 60-year oscillation is flipped (see Figure 4) relative to the quasi 60-year oscillation found in the global mid-latitude auroral catalog. Thus, the Hungarian catalog suggests that the pattern observed in the global mid-latitude auroral catalog from 1700 to 1900 could have been produced by the interference between the ~60 yr and the ~85 yr oscillations. However, the result found in the global mid-latitude aurora record is also partially supported by its strong peak in 1850, which is not visible in the Hungarian record (see Figure 1B). Further research is required to clarify this issue that may be due to data artifacts or some not yet understood physical property of auroral sightings. The quasi 60 yr oscillation observed in the Hungarian record positively correlates with the quasi 60 yr cycle observed in solar and climate activity (Scafetta, 2010, 2012a, 2012c, 2012d).

Climate records appear to be characterized by similar cycles as well. Quasi 60 yr and 80-90 yr oscillations are found in climate records throughout the Holocene (Chylek et al., 2012; Klyashtorin et al., 2009; Knudsen et al., 2011; Jevrejeva et al., 2008; Qian & Lu, 2010; Scafetta, 2010, 2012a, 2012b, 2012c), and are commonly associated with the Gleissberg solar cycle (Hoyt & Schatten, 1997). A quasi 45-year cycle is also found in some climate records (Chylek et al., 2012; Oliver, 2005).

We have used harmonic regression models to forecast auroral and solar irradiance activity, smoothed of the ~11 year solar cycle, during 1956-2050. The models predict the well-established solar cycle minimum in the 1970s, a solar cycle maximum in 2000-2002, and another significant solar minimum in the 2030s. This pattern would agree with the ACRIM TSI composite (Willson & Mordvinov, 2003; Scafetta & Willson, 2009) and with a harmonic irradiance solar model prediction made by Scafetta (2012c).

In conclusion, our findings support a hypothesis that the dynamical behavior of the Sun, the Heliosphere and the terrestrial magnetosphere are modulated by gravitational and magnetic planetary forces, resulting in specific resonance structures in the solar system (Brown, 1900; Fairbridge & Shirley, 1987; Hung, 2007; Jose, 1965; Landscheidt, 1988, 1999; Molchanov, 1968) that could influence solar activity by means of some synchronization mechanism (Pikovsky et al., 2001). This conforms to the planetary influence that has been observed on other stars as well (Cuntz et al., 2000; Gurdemir et al., 2012; Scharf, 2010; Shkolnik et al., 2003; Shkolnik et al., 2005; Wright et al., 2008). Additional investigations of these phenomena and preliminary physical mechanisms have been recently published (Scafetta, 2010, 2012a, 2012c, 2012d; Wolff & Patrone, 2010; Abreu et al., 2012; Tan & Cheng, 2012; see also the commentary by Charbonneau, 2013.).

**Acknowledgment:** The National Aeronautics and Space Administration supported Dr. Willson under contracts NNG004HZ42C at Columbia University, Subcontracts 1345042 and 1405003 at the Jet Propulsion Laboratory.

# Appendix

Figure 5 shows a reconstruction of the position of the Northern magnetic pole (NMP) since 200 A.D. from a number of sources. The figure shows that from 1550 to 1960 the NMP moved within a relatively small region in an almost cyclical way, and could not significantly alter that frequency of aurora records in Hungary.

Figure 6 shows a scanned picture of the pages in Nordlichtbeobachtungen in Ungarn by Rethly & Berkes (1963) collecting the Hungarian aurora record. This is justified by the fact that Rethly & Berkes (1963) is not easily available in Universities' libraries.

| year | n | year | n | year | n | year | n | year | n | year | n | year | n | year | n |
|---|---|---|---|---|---|---|---|---|---|---|---|---|---|---|---|
| 1523 | 1 | 1607 | 2 | 1681 | 1 | 1730 | 7 | 1778 | 1 | 1835 | 1 | 1880 | 1 | 1940 | 4 |
| 1556 | 1 | 1608 | 1 | 1684 | 1 | 1734 | 1 | 1779 | 4 | 1836 | 1 | 1882 | 1 | 1941 | 8 |
| 1557 | 1 | 1609 | 1 | 1692 | 1 | 1736 | 2 | 1780 | 5 | 1837 | 2 | 1892 | 1 | 1942 | 1 |
| 1579 | 2 | 1610 | 2 | 1704 | 1 | 1737 | 2 | 1781 | 3 | 1847 | 1 | 1898 | 2 | 1943 | 2 |
| 1580 | 3 | 1611 | 1 | 1713 | 2 | 1739 | 1 | 1782 | 1 | 1851 | 1 | 1903 | 1 | 1946 | 3 |
| 1583 | 1 | 1612 | 3 | 1716 | 1 | 1741 | 1 | 1783 | 4 | 1852 | 1 | 1905 | 1 | 1947 | 5 |
| 1591 | 2 | 1613 | 5 | 1719 | 1 | 1749 | 1 | 1784 | 2 | 1854 | 1 | 1908 | 1 | 1948 | 2 |
| 1592 | 1 | 1614 | 2 | 1720 | 1 | 1761 | 2 | 1785 | 1 | 1859 | 4 | 1909 | 1 | 1949 | 1 |
| 1593 | 2 | 1615 | 4 | 1721 | 2 | 1763 | 1 | 1786 | 1 | 1862 | 1 | 1910 | 1 | 1950 | 5 |
| 1599 | 3 | 1623 | 3 | 1724 | 1 | 1768 | 1 | 1787 | 8 | 1869 | 5 | 1917 | 3 | 1951 | 1 |
| 1600 | 1 | 1627 | 1 | 1725 | 3 | 1769 | 4 | 1788 | 5 | 1870 | 8 | 1920 | 1 | 1957 | 4 |
| 1602 | 1 | 1642 | 1 | 1726 | 1 | 1770 | 1 | 1789 | 1 | 1871 | 4 | 1926 | 2 | 1958 | 4 |
| 1604 | 2 | 1648 | 1 | 1727 | 1 | 1775 | 1 | 1806 | 3 | 1872 | 3 | 1938 | 2 | 1959 | 1 |
| 1605 | 10 | 1663 | 1 | 1728 | 1 | 1777 | 1 | 1831 | 4 | 1877 | 1 | 1939 | 2 | 1960 | 1 |

Table 1:

Number of auroral sightings per year: historical Hungarian aurora record, 1523-1960, (Rethly & Berkes, 1963). See also the Appendix.

|     | $h_1(t)$ | $h_2(t)$ | $h_3(t)$ | $h_4(t)$ | $f(t)$ |
|-----|----------|----------|----------|----------|--------|
| $A_1$ | 0.15 | -0.39 | -0.28 | 1.29 | 0.33 |
| $T_1$ | 1974.6 | 1989.6 | 1921.3 | 1957.2 | 1940.6 |
| $P_1$ | 197.6 | 84.1 | 56.9 | 40.3 | 176.8 |
| $A_2$ | 0.27 | 0.18 | -0.13 | -1.44 | 0.43 |
| $T_2$ | 1915.1 | 1940.0 | 1908.6 | 1958.9 | 1946.6 |
| $P_2$ | 154.6 | 72.1 | 51.7 | 40.6 | 81.9 |
| $A_3$ | -0.09 | 0.04 | 0.13 | 0.19 | 0.43 |
| $T_3$ | 1906.9 | 2073.3 | 1924.2 | 1960.1 | 1946.8 |
| $P_3$ | 119.2 | 119.5 | 64.3 | 45.1 | 55.8 |
| $A_4$ |  |  |  |  | 0.42 |
| $T_4$ |  |  |  |  | 1954.0 |
| $P_4$ |  |  |  |  | 43.6 |
| $C$ |  |  |  |  | 0.56 |

Table 2:

Regression coefficients found for Eq. (1) and Eq. (2) using the nonlinear least-squares (NLLS) Marquardt-Levenberg algorithm.

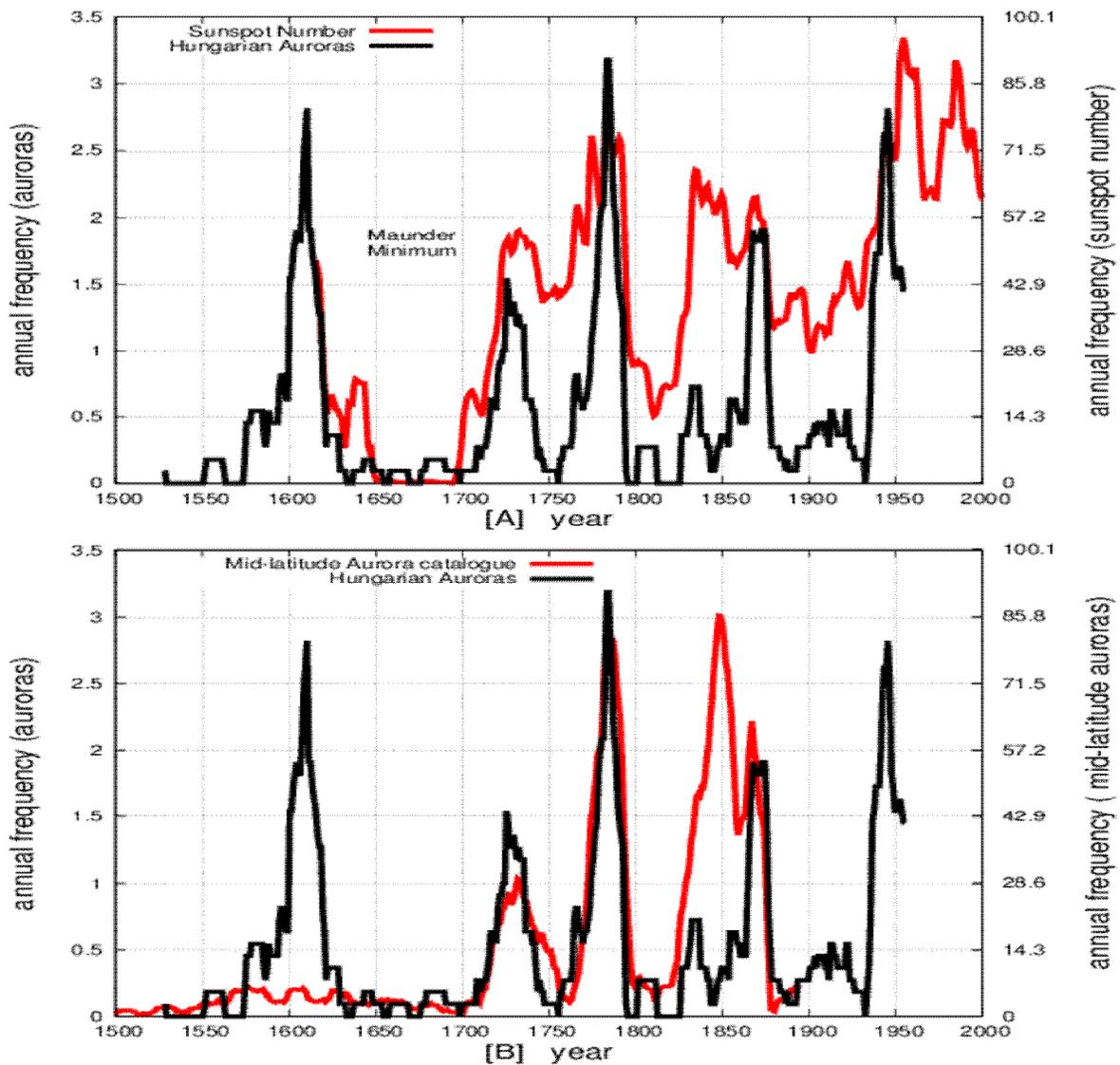

Figure 1:

[A] Hungarian auroral record (black) (Rethly & Berkes, 1963) (see Table 1) against the Sunspot number record (red): correlation coeff. r = 0.7, p < 0.0001. [B] Hungarian auroral record against the global mid-latitude aurora catalogue (red) (Křivský & Pejml, 1988): correlation coeff. r = 0.5, p < 0.0001. The data are processed with a 11-year moving average algorithm.

(Data from http://www.ngdc.noaa.gov/stp/aeronomy/aurorae.html and

http://www.ngdc.noaa.gov/stp/solar/ssndata.html).

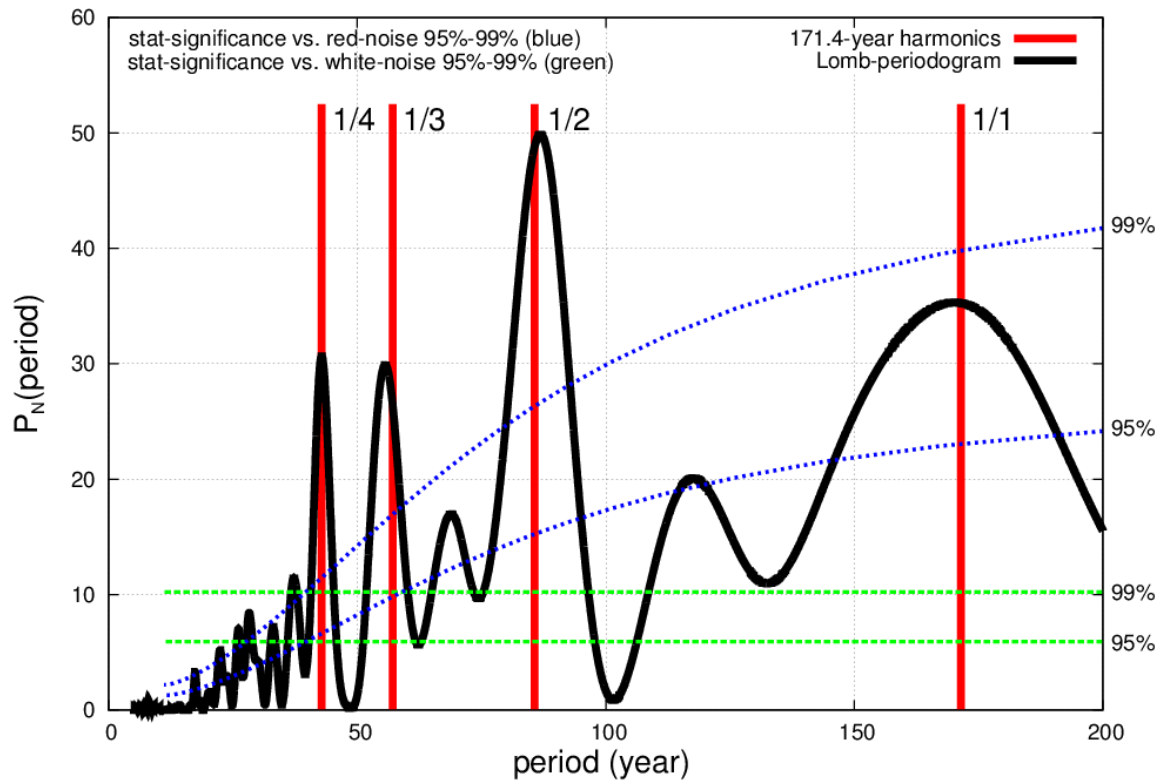

Figure 2:

Normalized Lomb periodogram of the Hungarian aurora record depicted in Figure 1 against the period. Four major power spectrum peaks are observed and form an almost perfect harmonic set. The red lines are the harmonic set at periods of about 42.85, 57.13, 85.7, 171.4 years. The later is Uranus/Neptune synodic period. Statistical significance at 95% and 99% of the frequency peaks is calculated against white and red noise backgrounds for the frequency range $0.005 < f < 0.091$ 1/yr (note that the implemented 11-year moving average algorithm dampens higher frequencies) (Horne and Baliunas, 1986; Press et al, 1997; Ghil M., et al., 2002).

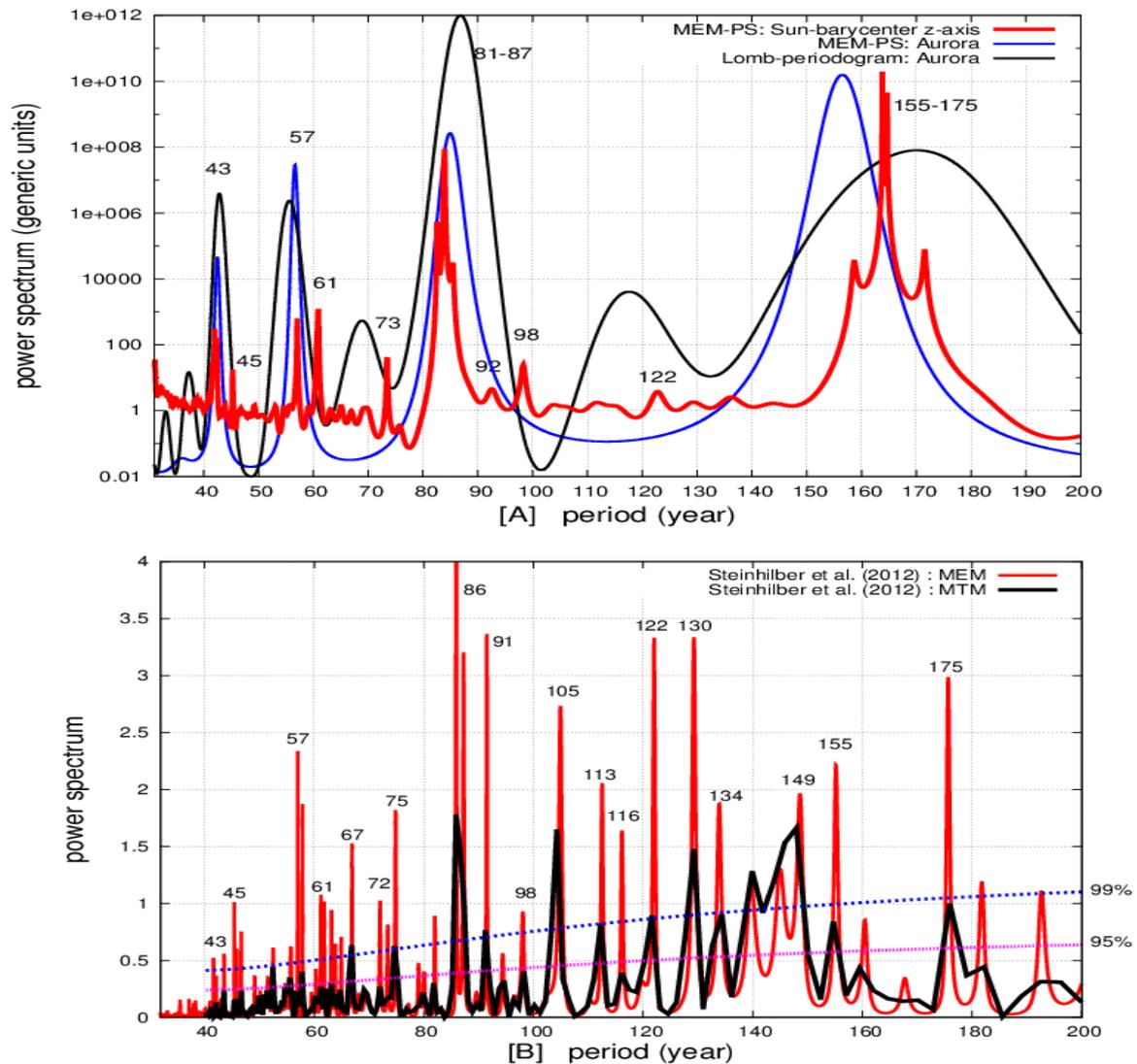

Figure 3:

[A] Power spectrum comparison between the z-axis coordinate of the Sun about the center of mass of the solar system (red) and the Hungarian auroral record (black, blue). Lomb-periodogram and Maximum Entropy Method (MEM) power spectra are used (Courtillot et al., 1977; Press et al., 1997). Note the four major common peaks. [B] Power spectrum evaluations using MEM and the multi taper method (MTM) of the cosmogenic records by Steinhilber, et al., (2012) covering 9,400 years of the Holocene against red noise backgrounds (significance at 95% and 99%).

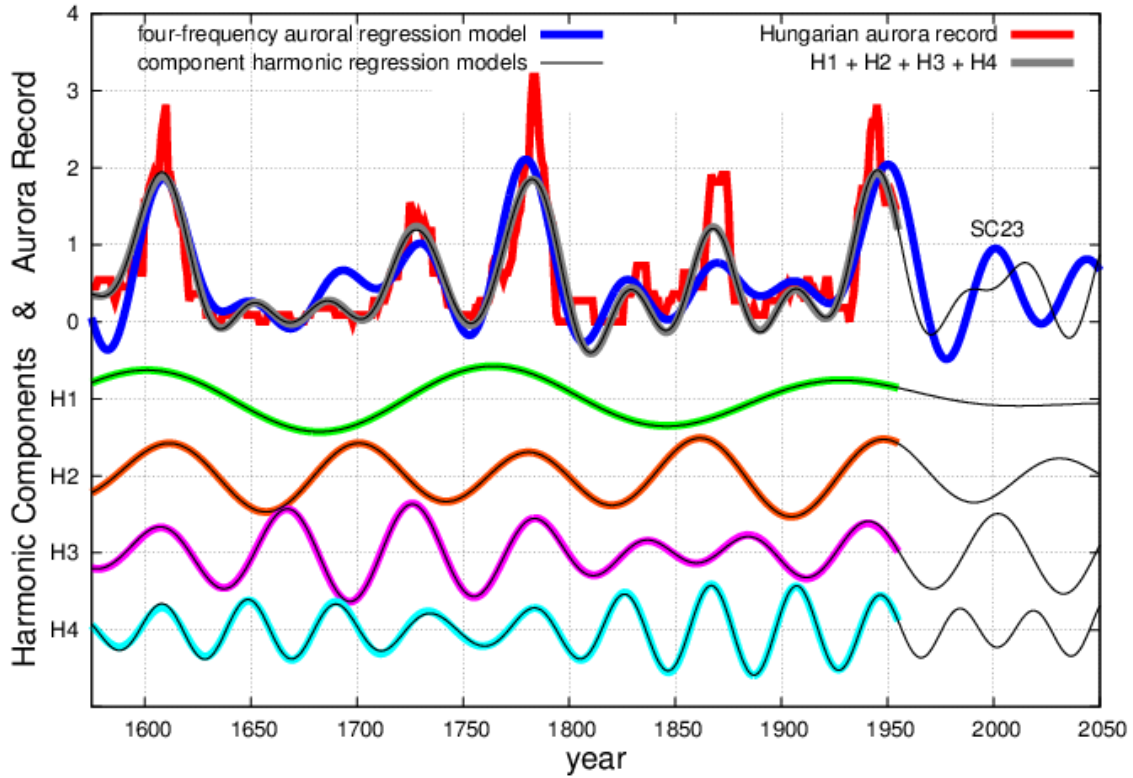

Figure 4:

Hungarian auroral record as in Figure 1 (red) and its Fourier band pass filter decomposition (colored curves in green, orange, purple and cyan): $H_1(x)$ captures the dynamical variability within the scale of 130-233 yr; $H_2(x)$, 74-130 yr; $H_3(x)$, 49-74 yr; $H_4(x)$, 35-49 yr. The gray curve is the sum of the four harmonic components. The black thin curves are regression harmonic models of the four components made of three frequencies each. The blue curve is a direct regression model of the auroral record that uses the four auroral frequencies highlighted in Figures 2 and 3 since 1600.

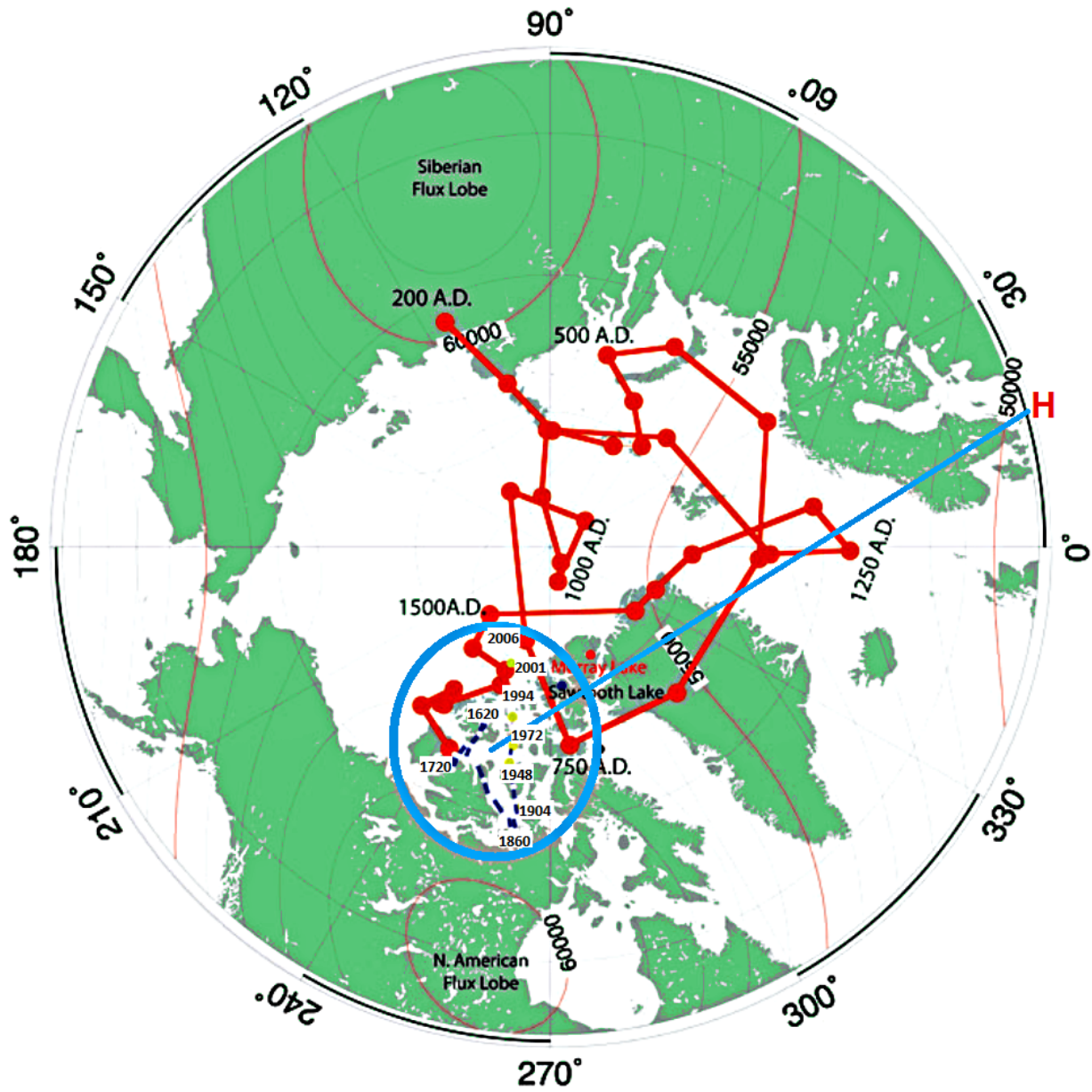

Figure 5:

A reconstruction of the position of the Northern Magnetic Pole (NMP) from 200 A.D. to present. Sorce: http://paleomag.coas.oregonstate.edu/research/projects/arctic/arcticfig2.html.
The cyan ring was added to indicate the area of interest in the present work. The red letter H approximately indicates the position of Hungary.

Tabelle 2

Die in Ungarn aufgezeichneten Nordlichter in den Jahren 1523–1960

### 16. Jahrhundert

| Nr. | Jahr | Tag | Monat | Stärke |
|---|---|---|---|---|
| 1. | 1523, | —  | —  | 2° |
| 2. | 1556, | 14. | I. | 2° |
| 3. | 1557, | 5. | XII. | 2° |
| 3a. | 1579, | 8. | II. | 2° |
| 3b. | 1579, | 9. | II. | 2° |
| 4. | 1580, | —  | IX. | 2° |
| 5. | 1580, | 12. | I. | 3° |
| 6. | 1580, | —  | XII. | 2° |
| 6a. | 1583, | 20. | VIII. | 1° |
| 7. | 1591, | aestas | | 2° |
| 8. | 1591,(20.) | XII. | | 2° |
| 9. | 1592, | 29. | XII. | 3° |

### 17. Jahrhundert

| Nr. | Jahr | Tag | Monat | Stärke |
|---|---|---|---|---|
| 15. | 1602, | 4. | XI. | 2° |
| 16. | 1604, 29. | IX. | | 3° |
| 17. | 1604, 24. | X. | | 4° |
| 18. | 1605, | 9. | II. | 2° |
| 19. | 1605, 15. | II. | | 2° |
| 20. | 1605, 16. | II. | | 2° |
| 21. | 1605, 17. | II. | | 2° |
| 22. | 1605, 18. | II. | | 2° |
| 23. | 1605, 19. | II. | | 2° |
| 24. | 1605, 22. | III. | | 2° |
| 25. | 1605, | 6. | VI. | 2° |
| 26. | 1605, 16. | XI. | | 2° |
| 27. | 1605, 17. | XII. | | 3° |
| 28. | 1607, 18. | I. | | 2° |
| 29. | 1607, 27. | XI. | | 2° |
| 30. | 1608, finis | II. | | 2° |
| 31. | 1609, | 8. | IX. | 3° |
| 32. | 1610, | —  | II. | 2° |
| 33. | 1610, 17. | XII. | | 2° |
| 34. | 1611, 27. | VIII. | | 3° |
| 35. | 1612, 10. | VII. | | 1° |
| 36. | 1612, | 4. | VIII. | 1° |
| 37. | 1612, 28. | VIII. | | 3° |
| 38. | 1613, 18. | VII. | | 1° |
| 39. | 1613, | 9. | XI. | 2° |
| 40. | 1613, 10. | XI. | | 2° |
| 41. | 1613, 12. | XI. | | 2° |
| 42. | 1613, 16. | XI. | | 2° |
| 43. | 1614, 20. | II. | | 1° |
| 44. | 1614, 23. | II. | | 2° |
| 44a. | 1615, | 7. | I. | 2° |
| 45. | 1615, 20. | IV. | | 1° |
| 46. | 1615, 17. | VI. | | 1° |
| 47. | 1615, | 2. | XI. | 1° |
| 48. | 1623, 13. | I. | | 3° |
| 49. | 1623, 17. | V. | | 3° |
| 50. | 1623, 11. | XI. | | 2° |
| 50a. | 1627, | —  | —  | 2° |
| 50b. | 1642, | 4. | I. | 2° |
| 51. | 1648, 11. | I. | | 1° |
| 52. | 1663, 25. | XI. | | 1° |
| 52a. | 1681, 10. | II. | | 3° |
| 52b. | 1684, 16. | X. | | 2° |
| 53. | 1692, 10. | II. | | 1° |

### 18. Jahrhundert

| Nr. | Jahr | Tag | Monat | Stärke |
|---|---|---|---|---|
| 54. | 1704, | 1. | VIII. | 2° |
| 55. | 1713, | —  | I. | 1° |
| 56. | 1713, | —  | III. | 1° |
| 57. | 1716, 17. | III. | | 2° |
| 58. | 1719, 23. | XII. | | 3° |
| 59. | 1720, | 3. | I. | 2° |
| 60. | 1721, 17. | II. | | 1° |
| 61. | 1721, 23. | II. | | 2° |
| 62. | 1724, 25. | XII. | | 1° |
| 63. | 1725, | 8. | I. | 1° |
| 64. | 1725, | 2. | I. | 1° |
| 65. | 1725, 17. | XII. | | 2° |
| 66. | 1726, 19. | X. | | 3° |
| 67. | 1727, 14. | III. | | 3° |
| 68. | 1728, 20. | I. | | 2° |
| 68a. | 1730, 12. | I. | | 1° |
| 69. | 1730, 14. | II. | | 1° |
| 70. | 1730, 15. | II. | | 4° |
| 71. | 1730, | 6. | III. | 1° |
| 72. | 1730, 21. | VI. | | 2° |
| 73. | 1730, | 7. | X. | 1° |
| 74. | 1730, | 3. | XI. | 3° |
| 75. | 1734, 25. | I. | | 2° |
| 76. | 1736, 15. | III. | | 3° |
| 77. | 1736, | 4. | V. | 2° |
| 77a. | 1737, | 1. | II. | 2° |
| 78. | 1737, 16. | XII. | | 3° |
| 78a. | 1739, 29. | III. | | 2° |
| 79. | 1744, 16. | XII. | | 2° |
| 79a. | 1749, | 3. | II. | 2° |
| 80. | 1761, 28. | II. | | 1° |
| 81. | 1761, 19. | XI. | | 1° |
| 82. | 1763, | —  | XI. | 3° |
| 83. | 1768, | 5. | XII. | 1° |
| 84. | 1769, 26. | II. | | 1° |
| 85. | 1769, 27. | II. | | 4° |
| 86. | 1769, 24. | X. | | 3° |
| 87. | 1769, 31. | X. | | 2° |
| 88. | 1770, 18. | X. | | 2° |
| 89. | 1770, | 3. | XI. | 3° |
| 90. | 1777, | 3. | XII. | 2° |
| 91. | 1778, 25. | II. | | 3° |
| 91a. | 1779, 10. | II. | | 1° |
| 91b. | 1779, 11. | II. | | 2° |
| 91c. | 1779, 13. | II. | | 3° |
| 91d. | 1779, | 9. | XI. | 2° |
| 92. | 1780, 29. | II. | | 1° |
| 93. | 1780, 28. | VII. | | 1° |
| 94. | 1780, 24. | XI. | | 1° |
| 95. | 1780, 25. | XI. | | 3° |
| 96. | 1780, 26. | XI. | | 1° |
| 97. | 1781, 30. | I. | | 1° |
| 98. | 1781, 16. | II. | | 1° |
| 99. | 1781, 27. | XII. | | 1° |
| 100. | 1782, | 8. | X. | 1° |
| 101. | 1783, 18. | III. | | 2° |
| 102. | 1783, 22. | III. | | 1° |
| 103. | 1783, 23. | III. | | 1° |
| 104. | 1783, 22. | X. | | 1° |
| 105. | 1784, 25. | VII. | | 1° |
| 106. | 1784, 15. | XI. | | 1° |
| 107. | 1785, | 9. | VIII. | 1° |
| 108. | 1786, 22. | III. | | 2° |
| 109. | 1787, 18. | III. | | 1° |
| 110. | 1787, 13. | V. | | 1° |
| 111. | 1787, 13. | VII. | | 2° |
| 112. | 1787, | 6. | X. | 2° |
| 113. | 1787, 10. | X. | | 1° |
| 114. | 1787, 13. | X. | | 1° |
| 115. | 1787, 24. | X. | | 1° |
| 116. | 1787, 31. | X. | | 2° |
| 117. | 1788, 11. | II. | | 1° |
| 118. | 1788, 15. | II. | | 1° |
| 119. | 1788, 19. | VIII. | | 1° |
| 120. | 1788, 21. | X. | | 1° |
| 121. | 1788, 22. | X. | | 1° |
| 122. | 1789, | 1. | IV. | 1° |

### 19. Jahrhundert

| Nr. | Jahr | Tag | Monat | Stärke |
|---|---|---|---|---|
| 123. | 1806, | 3. | XI. | 1° |
| 124. | 1806, | 4. | XI. | 1° |
| 125. | 1806, | 5. | XI. | 1° |
| 126. | 1831, | 7. | I. | 3° |
| 127. | 1831, 12. | I. | | 1° |
| 128. | 1831, 20. | I. | | 1° |
| 129. | 1831, 26. | IX. | | 1° |
| 130. | 1835, | 3. | XII. | 2° |
| 131. | 1836, 18. | X. | | 4° |
| 132. | 1837, 15. | III. | | 2° |
| 133. | 1837, | 5. | XI. | 1° |
| 133a. | 1847, 17. | XII. | | 3° |
| 134. | 1851, 26. | XII. | | 1° |
| 135. | 1852, 19. | II. | | 2° |
| 136. | 1854, | 2. | I. | 2° |
| 137. | 1859, 21. | IV. | | 2° |
| 138. | 1859, 28. | VIII. | | 2° |
| 139. | 1859, | 3. | IX. | 1° |
| 140. | 1859, 12. | X. | | 1° |
| 141. | 1862, 14. | XII. | | 1° |
| 142. | 1869, 15. | IV. | | 2° |
| 143. | 1869, 13. | V. | | 2° |
| 144. | 1869, 29. | V. | | 1° |
| 145. | 1869, 25. | X. | | 3° |
| 146. | 1869, 12. | XII. | | 1° |
| 147. | 1870, | 5. | IV. | 3° |
| 148. | 1870, 20. | V. | | 2° |
| 149. | 1870, 24. | IX. | | 2° |
| 150. | 1870, 14. | X. | | 1° |
| 151. | 1870, 23. | X. | | 1° |
| 152. | 1870, 24. | X. | | 4° |
| 153. | 1870, 25. | X. | | 4° |
| 154. | 1870, 19. | XI. | | 1° |
| 155. | 1871, 13. | II. | | 1° |
| 156. | 1871, | 2. | XI. | 2° |
| 157. | 1871, | 3. | XI. | 2° |
| 158. | 1871, | 5. | XI. | 2° |
| 159. | 1872, | 4. | II. | 4° |
| 160. | 1872, 11. | II. | | 2° |
| 161. | 1872, | 6. | VIII. | 1° |
| 162. | 1877, 29. | IX. | | 1° |
| 163. | 1880, 27. | X. | | 2° |
| 164. | 1882, | 2. | X. | 2° |
| 165. | 1892, 12. | VIII. | | 1° |
| 166. | 1898, | 9. | IX. | 2° |
| 167. | 1898, | 3. | XII. | 1° |

### 20. Jahrhundert

| Nr. | Jahr | Tag | Monat | Stärke |
|---|---|---|---|---|
| 168. | 1903, 31. | I. | | 1° |
| 169. | 1905, 15. | VI. | | 2° |
| 170. | 1908, 30. | VI. | | 1° |
| 171. | 1909, 28. | VII. | | 2° |
| 172. | 1910, | 3. | III. | 2° |
| 173. | 1917, 17. | XII. | | 1° |
| 174. | 1917, 18. | XII. | | 1° |
| 175. | 1917, 19. | XII. | | 3° |
| 176. | 1920, 22. | III. | | 2° |
| 177. | 1926, 26. | II. | | 1° |
| 178. | 1926, | 9. | III. | 1° |
| 179. | 1938, 25. | V. | | 4° |
| 180. | 1938, 12. | VII. | | 1° |
| 181. | 1939, 24. | II. | | 1° |
| 182. | 1939, 23. | IV. | | 1° |
| 183. | 1940, 23. | III. | | 1° |
| 184. | 1940, 24. | III. | | 4° |
| 185. | 1940, 27. | III. | | 1° |
| 186. | 1940, 17. | IV. | | 1° |
| 187. | 1941, 27. | I. | | 1° |
| 188. | 1941, 28. | I. | | 1° |
| 189. | 1941, 17. | II. | | 1° |
| 190. | 1941, | 1. | III. | 2° |
| 191. | 1941, | 2. | III. | 1° |
| 192. | 1941, 30. | III. | | 1° |
| 193. | 1941, 18. | IX. | | 3° |
| 194. | 1941, 22. | X. | | 2° |
| 195. | 1942, 23. | II. | | 1° |
| 196. | 1943, | 2. | II. | 1° |
| 197. | 1943, 22. | IX. | | 1° |
| 198. | 1946, 28. | III. | | 1° |
| 199. | 1946, 27. | VII. | | 2° |
| 200. | 1946, 28. | III. | | 1° |
| 201. | 1947, 17. | II. | | 1° |
| 202. | 1947, 12. | III. | | 1° |
| 203. | 1947, 17. | IV. | | 2° |
| 204. | 1947, 17. | VII. | | 1° |
| 205. | 1947, 16. | VIII. | | 1° |
| 206. | 1948, 20. | XI. | | 1° |
| 207. | 1948, 22. | XI. | | 1° |
| 208. | 1949, 26. | I. | | 2° |
| 209. | 1950, 21. | I. | | 1° |
| 210. | 1950, 24. | I. | | 2° |
| 211. | 1950, 25. | I. | | 3° |
| 212. | 1950, 26. | II. | | 2° |
| 213. | 1950, 20. | II. | | 2° |
| 214. | 1951, 28. | X. | | 2° |
| 215. | 1957, 21. | II. | | 1° |
| 216. | 1957, 26. | X. | | 1° |
| 217. | 1957, | 6. | XI. | 1° |
| 218. | 1957, 26. | XI. | | 2° |
| 219. | 1958, 10. | II. | | 3° |
| 220. | 1958, | 8. | VII. | 3° |
| 221. | 1958, | 4. | IX. | 4° |
| 222. | 1958, | 4. | XII. | 1° |
| 223. | 1959, | 4. | IX. | 1° |
| 224. | 1960, | 6. | X. | 3° |

Figure 6:

Scanned picture of the pages in Nordlichtbeobachtungen in Ungarn by Rethly & Berkes (1963) collecting the Hungarian aurora record (provided by Dr. Leif Svalgaard, personal communication).